# Unraveling Ultrafast Photoionization in Hexagonal Boron Nitride

Lianjie Xue[1,2,3*], Song Liu[2*], Yang Hang[4*], Adam M. Summers[1,3], Derrek J. Wilson[1,3,5], Xinya Wang[6], Pingping Chen[6], Thomas G. Folland[7], Jordan A. Hachtel[8], Hongyu Shi[1,3], Sajed Hosseini-Zavareh[1,3], Suprem R. Das[6,9], Shuting Lei[6], Zhuhua Zhang[4], Christopher M. Sorensen[1], Wanlin Guo[4], Joshua D. Caldwell[7], James H. Edgar[2], Cosmin I. Blaga[1,3], Carlos A. Trallero-Herrero[10]

[1]Department of Physics, Kansas State University, Manhattan, KS 66506, USA
[2]Tim Taylor Department of Chemical Engineering, Kansas State University, Manhattan, KS, 66506, USA
[3]J. R. Macdonald Laboratory, Department of Physics, Kansas State University, Manhattan, KS 66506, USA
[4]State Key Laboratory of Mechanics and Control of Mechanical Structures and Institute of Nano Science, Nanjing University of Aeronautics and Astronautics, Nanjing, 210016, P. R. China
[5]Advanced Laser Light Source and few-cycle Inc., Institut Nationale de la Recherche Scientifique, 1650 Boul. Lionel-Boulet, Varennes, QC, J3X 1P7, Canada
[6]Department of Industrial and Manufacturing Systems Engineering, Kansas State University, Manhattan, KS 66506, USA
[7]Department of Mechanical Engineering, Vanderbilt University, Nashville, Tennessee, 37212, USA
[8]Center for Nanophase Materials Science, Oak Ridge National Laboratory, Oak Ridge, 37830 USA
[9]Department of Electrical and Computer Engineering, Kansas State University, Manhattan, KS 66506, USA
[10]Department of Physics, University of Connecticut, Storrs, CT 06269, USA
*Equal author contribution
Emails: sor@phys.ksu.edu; chuwazhang@nuaa.edu.cn; wlguo@nuaa.edu.cn; edgarjh@ksu.edu; blaga@phys.ksu.edu; carlos.trallero@uconn.edu.

**The non-linear response of dielectrics to intense, ultrashort electric fields has been a sustained topic of interest for decades with one of its most important applications being femtosecond laser micro/nano-machining. More recently, renewed interests in strong field physics of solids were raised with the advent of mid-infrared femtosecond laser pulses, such as high-order harmonic generation, optical-field-induced currents, etc. All these processes are underpinned by photoionization (PI), namely the electron transfer from the valence to the conduction bands, on a time scale too short for phononic motion to be of relevance. Here, in hexagonal boron nitride, we reveal that the bandgap can be finely manipulated by femtosecond laser pulses as a function of field polarization direction with respect to the lattice, in addition to the field's intensity. It is the modification of bandgap that enables the ultrafast PI processes to take place in dielectrics. We further demonstrate the validity of the Keldysh theory in describing PI in dielectrics in the few TW/cm$^2$ regime.**

Laser-matter interactions in optically transparent materials have attracted intensive interests since the invention of the laser,[1,2] from the point of view of fundamental science and practical applications. One of the most important applications is femtosecond laser micro/nano-machining, which has revolutionized materials processing since its first demonstration in 1987, and was recognized to enable the finest ablation structures.[3-6] Renewed interest in the strong-field responses in solids occurred with the development of mid-infrared lasers pulses,[7,8] as dielectrics can withstand fields approaching their critical fields. Examples are, control of superconductivity,[9] high-order harmonic generation,[10,11] optical-field-induced current,[12] electronic structure and electric polarizability manipulation,[13] to name a few. Despite these significant developments experimentally, understanding of the underlying mechanisms is still in its early stage.[8] As ionization is the first step for all strong-field processes that are electron-initiated,[14] an explicit understanding of photoionization (PI) in dielectrics under ultrashort strong field is prerequisite to comprehensively understand all these strong-field phenomena. Conventionally, this is not achievable because avalanche ionization (AI) has been proposed compete with PI for dominating the femtosecond regime dynamics.[15-18] In this work, from a combination of nonlinear ionization characterization, density functional theory (DFT) and Keldysh[19] calculations we determine that the bandgap of hexagonal boron nitride (hBN) can be finely manipulated by femtosecond laser pulses not only as a function of laser pulse intensity which enables the PI to take place, but also as a function of field polarization direction. The Keldysh theory which laid the foundation of strong field physics treats atoms and solids in a similar manner,[19] it describes quantitatively well the electron dynamics in semiconductors at weak fields,[7,14,20] but provides immensely underestimated PI rates in dielectrics at intensities exceeding TW/cm$^2$.[8,21] In this work we demonstrate that when strong field induced bandgap modification is taken into account, the Keldysh theory is actually capable of describing PI in hBN, a dielectric, quantitatively.

The sample arrangement is shown in Fig. 1a. The laser pulses are normally incident onto the (0001) plane of a single-crystalline hBN, we measured the absolute transmission as a function of sample orientation with respect to the laser polarization and laser pulse intensity (Fig. 1b). The transmission as a function of pulse polarization angle displays a six-fold modulating pattern beyond the pulse peak intensity of 2.69 TW/cm$^2$ (Fresnel reflection incorporated), consistent with the six-fold symmetry of hBN. The experimental transmission curves (Fig. 1b) were then fitted with a sinusoidal function ($y = y_0 + A * \sin(\pi * \frac{x-x_c}{30})$) to extract the average transmission ($y_0$) and the modulation depths (amplitude of the modulation: $A * 2$) shown in Fig. 1c. The three-dimensional (3-D) band structure of hBN in Fig. 1d calculated via DFT shows that valence band maxima (VBM) and the conduction band minima (CBM) are located at the K and M points, respectively, consistent with previous theoretical and experimental determination of the indirect bandgap nature of hBN.[22,23] The lack of absorption at low intensities (Fig. 1c) is evidence of nonlinear absorption,[7] consistent with the much wider bandgap of hBN (5.95 $eV$) than the photon energy ($\sim 1.5\ eV$) of the 800 nm light. Thus, a nonlinear process is responsible for the electron transfer from the VBM to the CBM, with phonon assistance, as has been elucidated by Cassabois *et al*.[22] and depicted in Fig 1d.

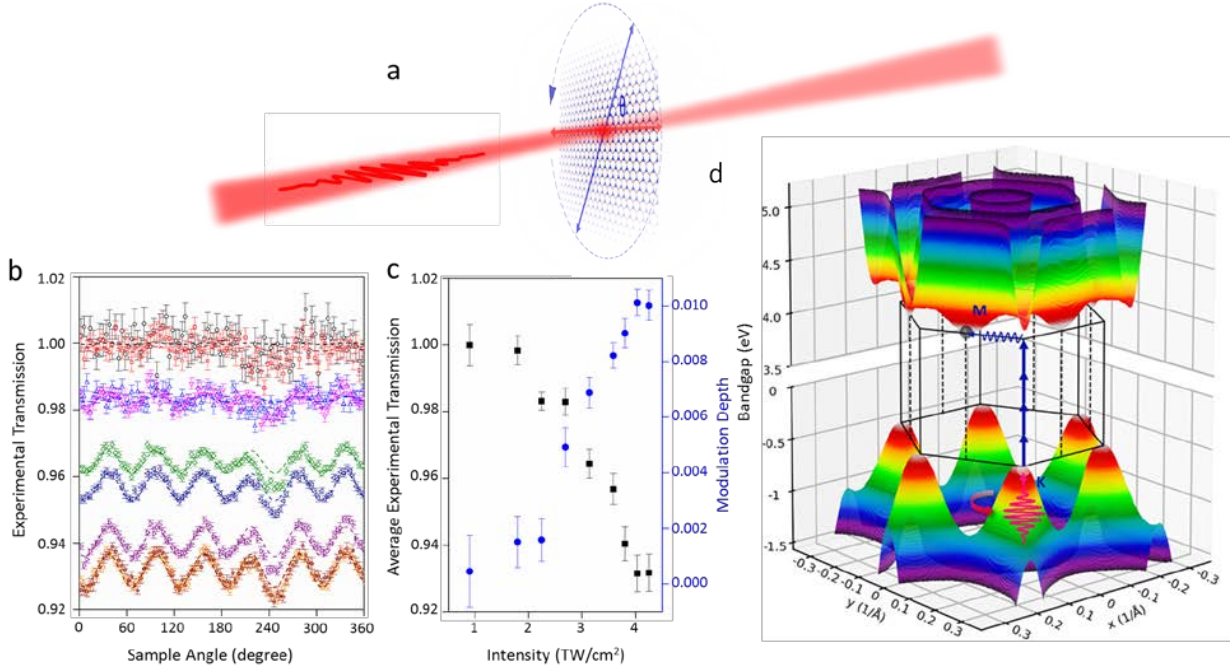

**Fig. 1 I Transmission measurement and electron excitation processes in hBN. a**, Schematic illustration of the transmission of laser pulses through an hBN single crystal which is rotated relative to pulse polarization. Note that during the measurement, the pulse polarization was rotated relative to the crystal orientation. The scheme is drawn in the opposite way for clarity. **b**, Transmission of 800 nm 33 fs linear laser pulses through a 6.25 μm h$^{11}$BN single crystal flake with normal incidence as a function of crystal angle at variant pulse intensities (black: 0.89 TW/cm$^2$, red: 1.79 TW/cm$^2$, blue: 2.23 TW/cm$^2$, magenta: 2.69 TW/cm$^2$, olive: 3.13 TW/cm$^2$, navy: 3.58 TW/cm$^2$, purple: 3.81 TW/cm$^2$, orange: 4.03 TW/cm$^2$, wine: 4.25 TW/cm$^2$, Fresnel reflection incorporated ). All the transmission curves have been lifted up according to the highest transmission to account for Fresnel reflection. **c,** Average transmission and modulation depths as a function of pulse intensity extracted from the transmission curves in Fig. 1b. **d**, 3-D representation of the electronic band structure of hBN calculated by DFT. An electronic excitation path across the bandgap from VBM to CBM via phonon assistance under ultrashort strong field, is marked by the blue arrows.

From the 3-D electronic band structure of hBN in Fig. 1d the effective masses of electrons and holes at the VBM and CBM, respectively, and the reduced effective mass of electron/hole pairs (m*) are calculated and plotted as a function of orientation angle in Fig. 2a.[7] Note that the zero angle degree in Fig. 2a as well as in the other angle dependent figures in this study corresponds to the x-axis direction of the 3-D electronic band structure of hBN in Fig. 1d. According to the hexagonal symmetry of hBN, there are two VBM and three CBM in the first Brillouin zone leading to six possible excitation paths, however Fig. 2a (bottom) depicts m* of only one of such paths for the sake of clarity. Full m* for all six paths as a function of crystal angle are plotted in Fig. S3e. In our Keldysh calculations, all six indirect transition rates were calculated and summed up to calculate the transmission. Without considering bandgap dynamics under strong field, the Keldysh theory is known to fail to describe PI processes in dielectrics, which can be verified to calculate the transmission of the 33 fs 800 nm pulses through hBN as a function of crystal angle at the experimental intensities, at hBN's intrinsic bandgap value of 5.95 $eV$. Calculated angle dependent transmission are plotted in Fig. 2b, with the average transmission and modulation depths as a function of pulse intensity displayed in Fig.2c. This theoretical approach failed to reproduce the experimental observations in three ways. First, the calculated PI rates are generally one order of magnitude lower than the experimental values, which appears as the calculated average transmission (~ 99.4 % at 4.25 TW/cm$^2$) being much closer to 100 % than the experimental value (~ 92.5 % at 4.25 TW/cm$^2$). This is expectable as the Keldysh theory is known for underestimating PI rates in dielectrics by up to 1-3 orders of magnitude.[24-27] Second, the theoretical modulation depth is nonmonotonic with pulse intensity unlike the experimental observation (compare Fig 2c with Fig 1c). Furthermore, we observe inverted 6-fold transmission modulation below and above 3.13 TW/cm$^2$ in the calculated transmission which is not observed in the experimental transmission. Third, the theoretical average transmission (Fig. 2c) decreases much slower than the experimentally measured one (Fig. 1c). The sharp decrease in transmission with field intensity observed in the experimental measurements will be attributed to a dynamic bandgap reduction under strong field in the later part of this study.

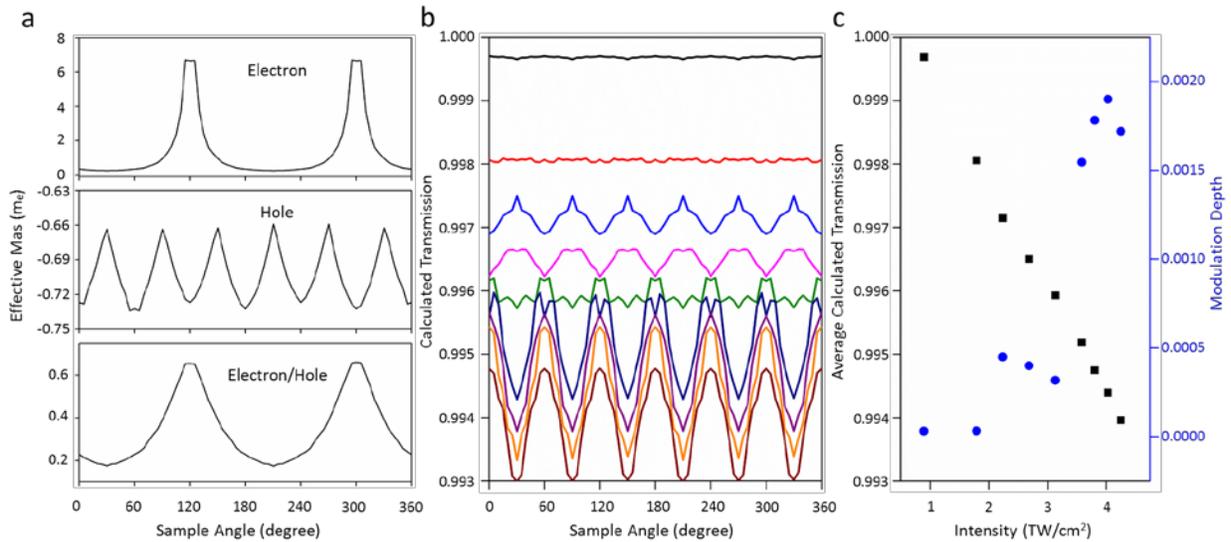

**Fig. 2 I Keldysh calculation of 800 nm 33 fs pulses through the 6.25 μm single crystal hBN at 5.95 eV. a**, Effective mass of electron (CBM), hole (VBM), and electron/hole pairs (m*) as a function of crystal angle. **b,** Keldysh calculated transmission of 800 nm 33 fs laser pulses through the 6.25 μm single crystal hBN flake as a function of crystal angle at the experimental intensities (black: 0.89 TW/cm$^2$, red: 1.79 TW/cm$^2$, blue: 2.23 TW/cm$^2$, magenta: 2.69 TW/cm$^2$, olive: 3.13 TW/cm$^2$, navy: 3.58 TW/cm$^2$, purple: 3.81 TW/cm$^2$, orange: 4.03 TW/cm$^2$, wine: 4.25 TW/cm$^2$, Fresnel reflection incorporated) at hBN's intrinsic bandgap of 5.95 eV. **c,** Average transmission and fitted modulation depth as a function of pulse intensity calculated from the transmission curves in Fig. 2b.

The deformation of quantum states due to the dynamic Stark effect in bound electrons is known to be of critical importance for multiphoton transitions[28] as it defines the very notion of strong-field,[29] and the bandgap of hBN has been calculated to decrease in presence of strong field.[30] With such insight, we examine the response of hBN's bandgap to transverse static electric field by performing DFT calculations,[30] and find that the bandgap of hBN is continuously reduced with increasing field strength and varies periodically with a six-fold pattern as a function of field orientation at high field strengths (Fig. 3a). From an atomistic point of view, the distortion of quantum states under an electric field is a second order effect with the field strength that depends on the density of states.[31] Considering the dramatic hBN bandgap reduction with increasing electric field strength, the six-fold bandgap variation around its *c* axis is a natural result from the relative orientation of in-plane electric field to the intrinsic electric polarization in hBN oriented along the B-N bonds. As such, the periodic variation of hBN bandgap under rotating femtosecond laser pulses provides a plausible cause of the six-fold experimental transmission modulations, which can be readily verified by the Keldysh calculation.

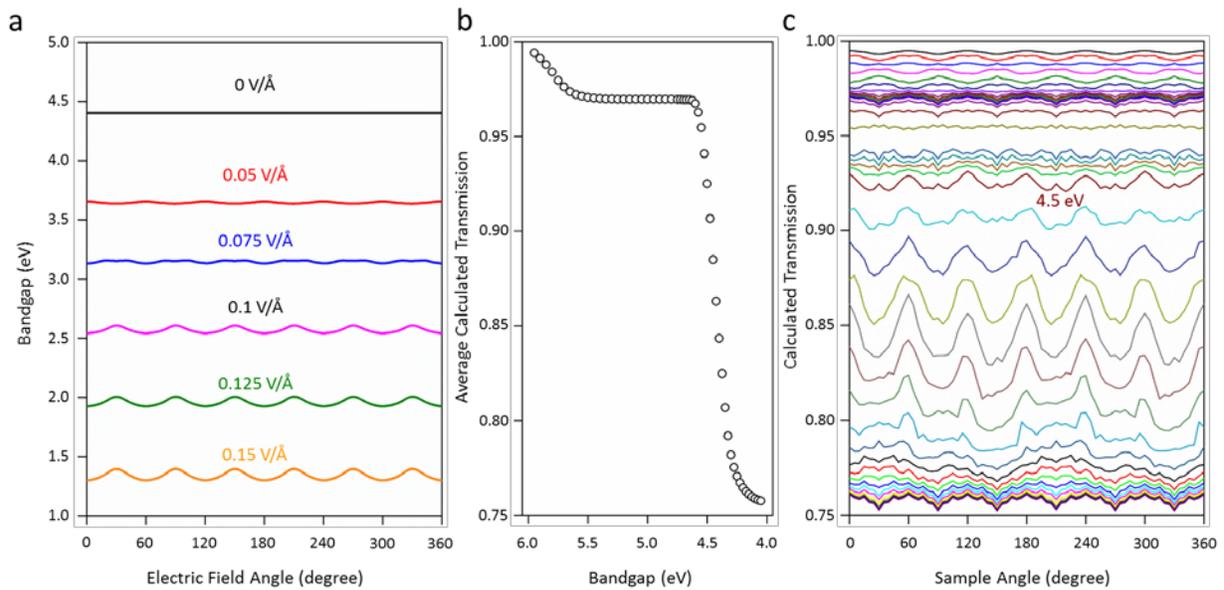

**Fig. 3 I DFT calculated hBN bandgap under transverse electric field and Bandgap dependent transmission calculation. a**, DFT calculated hBN bandgap under transverse electric field as a function of field direction at variant field strengths. **b,** Keldysh calculated average transmission of 800 nm 33 fs pulses through the 6.25 μm hBN single crystal as a function of bandgap at the pulse intensity of 4.25 TW/cm$^2$. **c,** Keldysh calculated transmission of 800 nm 33 fs laser pulses through 6.25 μm hBN single crystal as a function of crystal angle at decreasing bandgap. Each transmission curve in Fig. 3c corresponds to an average transmission point in Fig. 3b of the same height as these two figures are plotted with the same y-axis range.

From a PI point of view, dynamic resonances change the behavior of ionization as different levels go into and out of resonance, a behavior that is quantitatively included in the Keldysh model.[32] The influence of level dynamics under intense ionization fields has been established some time ago in atoms and are known as Freeman resonances.[33] To transpose such influence on PI in dielectrics the bandgap is proposed as a parameter depending on the field intensity and polarization orientation relative to the lattice. With these assumptions we calculated the transmission as a function of bandgap at the pulse peak intensity of 4.25 TW/cm$^2$ in Fig. 3b, which shows a strong plateau of transmission between 5.5 eV and 4.6 eV at which the gap is reduced by almost a full photon order, as is expected from dynamic resonances. Below 4.6 eV, the transmission declines sharply with decreasing bandgap. The trend of average transmission vs bandgap in Fig. 3b indicates hBN bandgap reduction along increasing pulse intensity, which would readily lead to a curved trend of average transmission vs pulse intensity as displayed by the experimental results. Given that the DFT method by nature underestimates bandgap, plus the difficulty of precisely converting pulse intensity into electric field strength, obtaining absolute bandgap values directly from DFT calculations for Keldysh calculation is infeasible. However, Fig. 1b, and Fig. 3a-c still render valuable clues upon which we propose an empirical bandgap vs pulse intensity model to carry out the Keldysh calculation. Fig. 3c provides that the anisotropic m* of hBN around its crystal axis also causes an anisotropic transmission, suppose the bandgap is isotropic around the crystal axis. In addition, from the DFT calculations we can extract that the overall bandgap shift follows an exponential law with intensity (field strength squared, see Fig. S2b), while the depth in bandgap modulation is linearly dependent on intensity (Fig. S2c).

The Keldysh parameter $\gamma = \omega \frac{\sqrt{E_g m^*}}{eE}$ where ω, $E_g$, $m^*$, and $E$ are the angular frequency of the light, the bandgap of the material, the reduced effective mass of carrier and the electric field strength associated with the laser pulses, respectively, determines the regime for PI. Multiphoton ionization (MPI) corresponds to γ > >1 and tunneling ionization (TI) for γ << 1. In our case, γ changes from γ ≈2 at the highest intensity to γ ≤8 at the lowest intensity. A great deal of insight can be learned by looking at the effective bandgap in the multiphoton regime, $\widetilde{\Delta}_{MPI} = E_g + \frac{eE^2}{4m\omega^2}$, where the second term in the sum corresponds to a ponderomotive energy or an equivalent Stark shift. While the ponderomotive energy is known to play a role in many processes such as HHG from solids, the parametrization for $\widetilde{\Delta}_{MPI}$ lacks bandgap dynamics which are also known to be extremely important[12] and an effect known for many decades. For all the calculations in this paper we use the full Keldysh theory (see Section 5 of the SI), but we ponder over $\widetilde{\Delta}_{MPI}$ for simplicity purposes and also because of $\widetilde{\Delta}_{MPI} \approx \widetilde{\Delta}$ in our regime. Thus, based on this and the ansatz provided by the DFT calculations we parametrize the effective bandgap in the following manner,

$$\tilde{\Delta}(I) = E_g + (eE^2)/(4m\omega^2) - \hbar\omega\big[\langle \tilde{x} \rangle - e^{-I/I_0}\big] + \delta E_g * \sin\big[(\theta - \theta_0) * N_\theta\big], \tag{1}$$

with $N_\theta$ being the symmetry order for the crystal ($N_\theta = 6$ for hBN), $I_0$ an effective threshold intensity at which the band gap dynamics dominates the transmission, and $\langle \tilde{x} \rangle$ is the effective multiphoton order $\langle \tilde{x} \rangle = \frac{\tilde{\Delta}_{MPI}(I)}{\hbar\omega}$. It is a well-established fact that one strong effect in gap closure is from the increase of carrier population in the conduction band.[34] However, as stated in Ref.[35] such effects should be negligible during the duration of our pulses. And from our DFT calculation, the strong electric field associated with the intense pulses reduces the bandgap effectively. To our surprise, we need to include an exponential factor in $\tilde{\Delta}(I)$ proposed in Ref.[36] which is a correction to the Wolf model for gap closure with the number of carriers in the conduction band. The last term is a new sinusoidal term due to the influence of field orientation on the bandgap dynamics as suggested by the results in Fig. 3a from DFT.

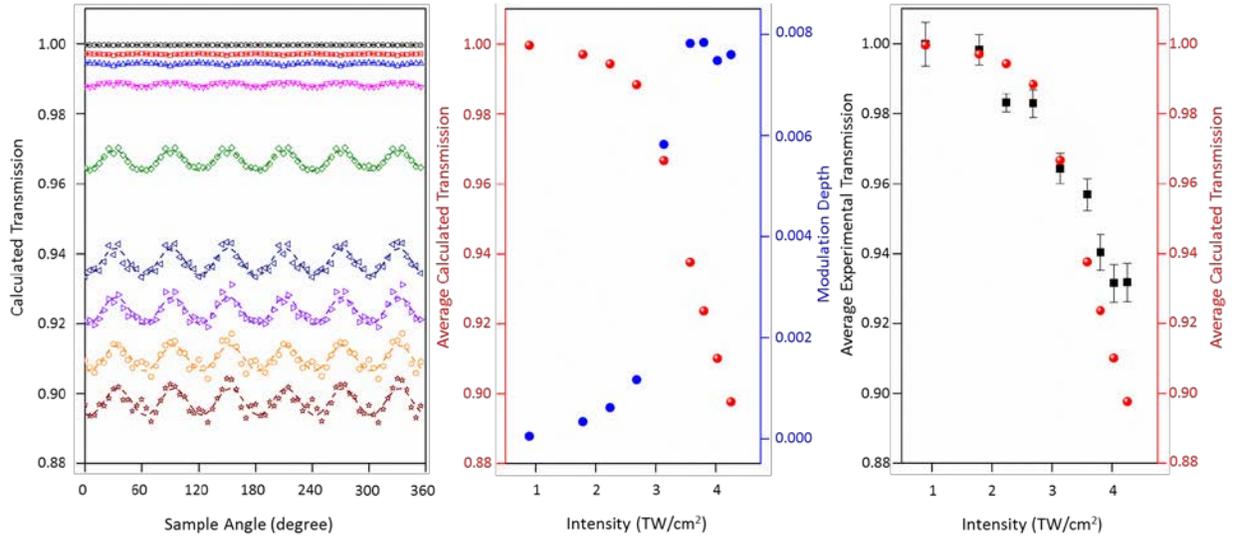

**Fig. 4 | Keldysh calculated transmission of 800 nm 33 fs laser pulses through 6.25 μm hBN single crystal under modeled bandgaps. a,** Transmission as a function of crystal angle at experimental intensities the same as in Fig. 1a. **b**, Average transmission and modulation depth as a function of pulse intensity extracted from Fig. 4a. **c,** Comparison between calculated and experimental average transmission as a function of pulse intensity.

Keldysh calculation of the transmission under the experimental conditions were carried out using parametrized bandgap $\tilde{\Delta}(I)$ described above (effective bandgap parametrization details are provided in section 7 of the SI). Six-fold modulating is observed in the newly calculated transmission in Fig. 4a with average transmission and transmission modulation depths depicted in Fig. 4b. Four consistencies between the newly calculated and the experimental transmission are summarized as below, which corroborates the quantitative recovery of the experimental results by the Keldysh calculation. First, the calculated and experimental average transmission shown in Fig. 4c are in very good agreement. Second, the monotonic transmission modulation depth increase with pulse intensity exhibited in the experimental transmission in Fig. 1b is observed in the calculated transmission in Fig. 4b. Third, the modulation saturation in the experimental transmission is also seen in the calculated transmission, a result from the interplay between the modulation in the bandgap of hBN and the anisotropy of m* around the hBN's crystal axis which cause counteractive transmission variation (compare Fig. 3a vs Fig. 3c). Fourth, the trend of calculated average transmission as a function of pulse intensity exhibits a clear curved pattern, similar to the experimental observation. This is directly associated with bandgap reduction under strong field, and can be rationally predicted by Fig. 3b. In fact the trend of curved shape in the average transmission vs pulse intensity is ubiquitously associated with wide bandgap crystals.[37] We also performed transmission measurements of ZnO (3.3 eV), CVD diamond (6.0 eV), MgO (7.8 eV), and CaF$_2$ (11.8 eV) with the same 800 nm femtosecond laser pulses. Not surprisingly, the curved pattern of average transmission vs pulse intensity becomes more obvious with wider bandgap crystals, but not observed in the narrow bandgap ZnO as shown in Fig. S8. Fig. S9 shows periodically modulating transmission curves around crystal axis in the three wide bandgap crystals, but not in ZnO. However, with longer wavelength and stronger field strength, ZnO also exhibited four-fold modulating HHG as a direct consequence of modulating PI around its crystal axis.[10] So we assert that a curved shape of average transmission vs pulse intensity is a feature of bandgap reduction under strong field.

The quantitative recovery of the experimental transmission by the Keldysh calculation validates the empirical hBN bandgap model, confirming that the bandgap of hBN can be finely manipulated by femtosecond laser pulses as a function of pulse intensity and polarization. Bandgap reduction under strong field inside a dielectric has been reported earlier,[12] but this is for the first time to incorporate such bandgap reduction into Keldysh calculation. The Keldysh theory is known to provide quantitative descriptions of PI in semiconductors in weak field, but underestimates PI rates immensely in dielectrics in the strong field regime. As a result, the physics responsible for the fundamental difference between PI in semiconductors and in dielectrics remains unresolved for decades,[14] and massive efforts have been devoted to modifying the Keldysh model to be more generalized.[14,20,21] We show here that by taking strong-field induced bandgap reduction into account, the Keldysh model without further modification is actually capable of providing a quantitative description of PI inside hBN. And it is just the reduction of bandgap in dielectrics under strong fields, which has been overlooked in past Keldysh calculations,[35] that dictates differing PI in semiconductors from that in dielectrics. An unprecedented understanding of PI inside a dielectric is achieved for the first time, which is beneficial to understanding all other PI involved strong field processes in solids.

The immense underestimation of PI in dielectrics by past Keldysh calculations has made researchers be reluctant to ascribe PI as the principal driving force for electron excitation in femtosecond laser micro/nano-machining,[6,38-40] which leads to the dilemma that the fundamental mechanism of ultrafast laser irradiation of matter remains unclarified so far.[41] However, from the quantitative Keldysh calculation of PI in hBN in this study, it is convincing that PI dominates the electron transfer to enable ultrafine material processing by femtosecond laser pulses. The short dwelling of femtosecond laser pulses and the nonlinear nature of PI, especially at the optical critical intensity, is assuring for the high deterministic feature of femtosecond laser micro/nano-machining.

Additionally, this is also the first report of the periodic modification of a dielectric's bandgap under rotating laser pulses, which accordingly causes a periodic PI in unison with its crystal symmetry. This provides a strong plausibility of the cause of the anisotropic HHG from optically isotropic crystals, which was reported recently[11,42] but not interpreted from the perspective of strong-field induced bandgap dynamics that indeed exist.

In conclusion, the PI inside an hBN single crystal is thoroughly investigated by a combination of experimental nonlinear ionization characterization, DFT and Keldysh calculations. We have determined that the bandgap of hBN could be finely manipulated by femtosecond laser pulses as a function of pulse intensity and polarization direction. While the former facilitates the PI to take place, the latter consists of a new order parameter $\delta E_g(\theta)$ in the modification of bandgap of dielectrics. Taking into account this new order parameter enables an empirical bandgap modification model, based on which the Keldysh theory can describe PI quantitatively in a dielectric for the first time. This provides the physics responsible for the fundamental difference between PI in semiconductors and in dielectrics which remains unresolved for decades. The unraveling of the PI inside hBN provides a valuable insight to understand all PI involved strong-field processes in solids. We hope our results will stimulate experimentalists and theoreticians to pursue further insights into strong-field physics of solids.

## Methods Summary

**Experimental transmission measurement.** The hBN sample characterized in this study is a monoisotropic B-11 enriched single crystal h$^{11}$BN which was synthesized by a metal solution precipitation method.[43] The single crystal domain of an hBN flake was mounted over an aperture, 300 µm in diameter, in an optical flat. Linearly polarized 800 nm 33 fs laser pulses with repetition rate of 1 kHz from a chirp pulse amplification based, multipass, Ti-Sapphire system (KM Laboratory, modified Red Dragon) was used to study the PI processes inside the h$^{11}$BN single crystal. The polarization alignment of the beam pulse was controlled by a half-wave plate (HWP). The beam was focused by a 500 mm focusing lens to an approximately 35 µm 1/e$^2$ radius spot. Further details of the experiments are provided in Section 1 of the SI. The thickness of the hBN is 6.25 µm, as determined by infrared spectroscopy method in Section 4 of the SI.

**DFT calculation.** The first-principles calculations were carried out within the framework of density functional theory, based on projector-augmented wave method[44] with generalized gradient approximation (GGA) of the Perdew–Burke–Ernzerhof (PBE) functional,[45] as implemented in the VASP package.[46] A kinetic energy cutoff of 400 eV was chosen for the plane-wave expansion. For the bulk h-BN, a 25 × 25 × 9 Monkhorst–Pack k-point mesh was initially used to sample the Brillouin zone for the self-consistent calculation and a denser mesh with up to 101 *101 *1 k-points was adapted for the three dimensional electronic structure calculation on the Kz=0 plane. One-dimensionally stacked circular H-terminated h-BN flakes were applied for calculating hBN bandgap change under electric field. The vacuum region was set to at least 15 Å in all directions within the plane to isolate neighboring periodic images. The flake consists of periodic layers of h-BN staked in the favorable AA'A mode, and the interlayer distance is set to the experimentally measured 0.34 nm. We considered two different flakes with radius R = 10.0 and 11.3 Å, respectively. The positions of all atoms were relaxed using the conjugate-gradient method until the force on each atom is less than 0.01 eV/Å. The in-plane external electric field was simulated by using a periodic saw-tooth-type potential across the BN flakes. This electric field was rotated by five degrees stepwise and we totally take 72 shots. Both flakes behave qualitatively the same with a six fold rotation symmetry in the band gap.

**Keldysh calculation.** A brief introduction of the Keldysh theory and the formulas for the Keldysh calculation are provided in Section 5 of the SI.

**Author Contributions** C.M.S., J.H.E., S. L., S.R.D, C.I.B and C.T.H. supervised the study. L.X., A.M.S., D.J.W., X.W., S.H.Z. and P.C. conceived and performed the measurements. S.L. and L.X. grew the hBN crystal. S.L., A.M.S., L.X. and X.W. participated in mounting the sample. T.G.F. determined the sample thickness. J.A.H. plotted the 3-d hBN band structure. Y.H., Z.Z. and W.G. performed the first-principles calculations. L.X., A.M.S. and C.T.H. analyzed and interpreted the experimental data. All authors discussed the results and contributed to the final manuscript.

**Acknowledgements** James R. Macdonald Laboratory, supported by U.S. Department of Energy (DOE), Office of Basic Energy Sciences, Chemical Sciences, Geosciences, and Biosciences Division, (DE-FG02-86ER13491); A.M.S. was supported by the Department of Defense (DoD) through the National Defense Science & Engineering Graduate Fellowship; D.J.W. was supported by the NSF Graduate Research Fellowship under Grant No. DGE-1247193; NSF EPSCoR IRR Track II Nebraska-Kansas Collaborative Research Award No. 1430493; hBN crystal growth was supported by NSF grant No. CMMI-1538127; L.X. was supported by the joint Dean's fund of college of Engineering and Arts and Science of Kansas State University; XW was supported by NSF grants CMMI-1537846 and CMMI-1903740. Some research was conducted at the Center for Nanophase Materials Sciences at Oak Ridge National Laboratory, which is a Department of Energy (DOE) Office of Science User Facility, through a user proposal. The work at NUUA is supported by the National Key Research and Development Program of China (2019YFA0705400), National Natural Science Foundation of China (11772153, 22073048), the Natural Science Foundation of Jiangsu Province (BK20190018). Part of this work was supported by Office of Naval Research, Directed Energy Ultra-Short Pulse Laser Division grant N00014-19-1-2339.

**References**

1. Du, D., Liu, X., Korn, G., Squier, J. & Mourou, G. Laser‐induced breakdown by impact ionization in SiO$_2$ with pulse widths from 7 ns to 150 fs. *Appl. Phys. Lett.* **64**, 3071-3073 (1994).
2. Bloembergen, N. Laser-Induced Electric Breakdown in Solids. *IEEE J. Quantum Electron.* **QE-10**, 375 (1974).
3. Malinauskas, M. *et al.* Ultrafast alser processing of materials: from science to industry. *Light Sci. Appl.* **5**, e16133 (2016).
4. Jiang, L., Wang, A.-D., Li, B., Cui, T.-H. & Lu, Y.-F. Electrons dynamics control by shaping femtosecond laser pulses in micro/nanofabrication: modeling, method, measurement and application. *Light Sci. Appl.* **7**, 17134 (2018).
5. Srinivasan, R., Sutcliffe, E. & Braren, B. Ablation and etching of polymethylmethacrylate by very short (160 fs) ultraviolet (308 nm) laser pulses. *Appl. Phys. Lett.* **51**, 1285-1287 (1987).
6. Sugioka, K. & Cheng, Y. Ultrafast lasers-reliable tools for advanced materials processing. *Light. Sci. Appl.* **3**, 1-12 (2014).
7. Golin, S. M. *et al.* Strong field processess inside gallium arsenide. *J. Phys. B: At. Mol. Opt. Phys.* **47**, 204025-204029 (2014).
8. Ghimire, S. *et al.* Strong-filed and attosecond physics in solids. *J. Phys. B: At. Mol. Opt. Phys.* **47**, 204030(204010) (2014).
9. Cavalleri, A. Photo-induced superconductivity. *Contemp. Phys.* **59**, 31-46 (2018).
10. Ghimire, S. *et al.* Observation of high-order harmonic generation in a bulck crystal. *Nat. Phys.* **7**, 138-141 (2011).
11. You, Y. S., Reis, D. A. & Ghimire, S. Anisotropic High-Harmonic Generation in Bulck Crystals. *Nat. Phys.* **13**, 345-350 (2017).
12. Schiffrin, A. *et al.* Optical-field-induced current in dielectrics. *Nature* **493**, 70-74 (2013).
13. Schultze, M. *et al.* Controlling dielectrics with the electric field of light. *Nature* **493**, 75-78 (2013).
14. McDonald, C. R., Vampa, G., Corkum, P. B. & Braec, T. Intense-laser solid state physics: unraveling the difference between semiconductors and dielectrics. *Phys. Rev. Lett.* **118**, 173601-173605 (2017).
15. Rajeev, P. P., Gertsvolf, M., Corkum, P. B. & Rayner, D. M. Field dependent avalanche ionization rates in dielectrics. *Phys. Rev. Lett.* **102**, 083001-083004 (2009).
16. Dachraoui, H., Oberer, C. & Heinzmann, U. Femtosecond crystallographic experimental in wide-bandgap LiF crystal. *Opt. Exp.* **19**, 2797-2804 (2011).
17. Quere, F., Guizard, S. & Martin, P. Time-resolved study of laser induced breakdown in dielectrics. *Europhys. Lett.* **56**, 138-144 (2001).
18. Wu, A. Q., Chowdhury, I. H. & Xu, X. Femtoscond laser ablation in fused silica: numerical and experimental investigation. *Phys. Rev. B* **72**, 085128-085127 (2005).
19. Keldysh, L. V. Ionization in the field of a strong electromagnetic wave. *So. Phys. -JETP* **20**, 1307-1314 (1965).
20. Gruzdev, V. E. Photoionization rate in wide band-gap crystals. *Phys. Rev. B* **75**, 205106 (2007).
21. Zhokhov, P. A. & Zheltikov, A. M. Field-cycle-resolved photoionization in solids. *Phys. Rev. Lett.* **113**, 133903(133905) (2014).
22. Cassabois, G., Valvin, P. & Gil, B. Hexagonal boron nitride is an indirect bandgap semiconductor. *Nat. Photonics* **10**, 262-266 (2016).

23  Arnaud, B., Lebegue, S., Rabiller, P. & Alouani, M. Huge excitonic effects in layered hexagonal boron nitride. *Phys. Rev. Lett.* **96**, 026402-026404 (2006).
24  Vaidyanathan, A., Walker, T., Guenther, A. H., Mitra, S. S. & Narducci, L. M. Two-photon absorption in several direct-gap crystals. *Phys. Rev. B* **21**, 743-748 (1980).
25  Schaffer, C. B., Brodeur, A. & Mazur, E. Laser-induced breakddown and damages in bulk transparent materials induced by tightly focused femtosecond laser pulses. *Meas. Sci. Technol.* **12**, 1784-1794 (2001).
26  Lenzner, M. *et al.* Femtosecond optical breakdown in dielectrics. *Phys. Rev. Lett.* **80**, 4076-4079 (1998).
27  Sudrie, L. *et al.* Femtosecond laser-induced damage and filamentary propagation in fused silica. *Phys. Rev. Lett.* **89**, 186601-186604 (2002).
28  Trallero-Herrero, C. A., Cohen, J. L. & Weinacht, T. Strong field atomic phase matching. *Phys. Rev. Lett.* **96**, 063603(063604) (2006).
29  Trallero-Herrero, C. A. & Weinacht, T. C. Transition from weak- to strong-field coherent control. *Phys. Rev. A* **75**, 063401(063408) (2007).
30  Zhang, Z. & Guo, W. Energy-gap modulation of BN ribbons by transverse electric fields: first princiles calculations. *Phys. Rev. B* **77**, 075403-075405 (2008).
31  Landau, L. D. & Lifshitz, E. M. *Quantum mechanics non-relativistic theory*. 3rd edn, (1977).
32  L., C. S., Rolland, C., Corkum, P. B. & Kelly, P. Multiphoton ionization of Xe and Kr with intense 0.62-μm femtosecond pulses. *Phys. Rev. Lett.* **61**, 153-156 (1988).
33  Freeman, R. R. *et al.* Above-threshold ionization with subpicosecond laser pulses. *Phys. Rev. Lett.* **59**, 1092-1095 (1987).
34  Wolff, P. A. Theory of the Band Structure of Very Degerate Semiconductors. *Phys. Rev.* **126**, 405-412 (1962).
35  Gruzdev, V. & Sergaeva, O. Ultrafast modification of band structure of wide-band-gap solids by ultrafast pulses of laser-driven electron oscillations. *Phys. Rev. B* **98**, 115202(115215) (2018).
36  Prabhu, S. S. & Vengurlekar, A. S. Dynamics of the pump-probe reflectivity spectra in GaAs and GaN. *J. Appl. Phys.* **95**, 7803-7812 (2004).
37  Grojo, D. *et al.* Long-wavelength multiphoton ionization inside band-gap solids. *Phys. Rev. B* **88**, 195135 (2013).
38  Phillips, K. C., Gandhi, H. H., Mazur, E. & Sundara, S. K. Ultrafast laser processing of materials: a review. *Adv. Opt. Photonics* **7**, 686-712 (2015).
39  Joglekar, A. P., Liu, H. H., Meyhofer, E., Mourou, G. & Hunt, A. J. Optics at critical intensity: applications to nanomorphing. *Proc. Natl. Acad. Sci. U. S. A.* **101**, 5856-5861 (2004).
40  Gattass, R. R. & Mazur, E. Femtosecond laser micromachining in transparent materials. *Nat. Photonics* **2**, 219-225 (2008).
41  Sugioka, K. Progress in ultrafast laser processing and future prospects. *Nanophotonics* **6**, 393-413 (2017).
42  Vampa, G. *et al.* Generation of high harmonics from silicon. *Preprint at* *http://arXiv.org/abs/1605.06345* (2016).
43  Liu, S. *et al.* Single crystal growth of millimeter-sized monoisotopic hexagonal boron nitride. *Chem. Mater.* **30**, 6222-6225 (2018).
44  Bolchl, P. Projector augmented wave method. *Phys. Rev. B* **50**, 17953-17979 (1994).
45  Perdew, J. P., Burke, K. & Ernzerhof, M. Generalized gradient approximation made simple. *Phys. Rev. Lett.* **77**, 3865-3868 (1996).
46  Kresse, G. Efficient iterative schemes for ab initio total-energy calculations using a plane-wave basis set. *Phys. Rev. B* **54**, 11169-11186 (1996).